\documentstyle{article}\begin{document}
\title{Relativistic diffusion  with friction on a pseudoriemannian  manifold }
\author{ Z. Haba\\
Institute of Theoretical Physics, University of Wroclaw,\\ 50-204
Wroclaw, Plac Maxa Borna 9, Poland}
\date{\today}\maketitle
\begin{abstract}
We study a relativistic  diffusion equation on the Riemannian phase
space defined by Franchi and Le Jan. We discuss stochastic Ito
(Langevin) differential equations  as a perturbation by noise of the
geodesic equation. We show that the expectation value of the angular
momentum and the energy grow exponentially fast.  We discuss drifts
leading to an equilibrium. As an example we consider a particle in
de Sitter universe. It is shown that the relativistic diffusion of
momentum in de Sitter space is the same as the relativistic
diffusion on the Minkowski mass-shell with the temperature
proportional to the de Sitter radius. We study a diffusion process
with a drift corresponding to the J\"uttner or quantum equilibrium
distributions. We show that such a diffusion has a bounded
expectation value of angular momentum and energy. The energy and the
angular momentum tend exponentially fast to their equilibrium
values.
 \end{abstract}

 \section{Introduction}
In this paper we discuss relativistic dynamics of a particle in
general relativity which moves in a medium (gas) of other particles.
We assume that the interaction with a gas consists of frequent
elastic collisions which in a Markov limit (no memory) can be
approximated by a diffusion process. In the framework of the special
theory of relativity the diffusion problem has been formulated and
solved a long time ago by Schay \cite{schay} and Dudley
\cite{dudley} under the assumption that the diffusion is evolving in
the proper time on the phase space and  preserves the mass $p^{2}$.
The diffusion problem on a general pseudoriemannian manifold has
been formulated recently in a geometric framework of diffusions on
fiber bundles by Franchi and Le Jan \cite{lejan}. The theory has
been further developed and applied in \cite{fran}
\cite{franha}\cite{bai}\cite{bai2} (for a general coordinate
dependent discussion of relativistic diffusions on the
pseudoriemannian manifolds see \cite{debbasch}).

We follow the method of our earlier papers
\cite{haba}\cite{habampa}\cite{habajpa} to define the relativistic
Brownian motion on the submanifold of fixed $p^{2}$ of the
Riemannian phase space. With our choice of coordinates the generator
of the process and the stochastic equations coincide with those of
Franchi and Le Jan \cite{lejan}. The diffusion process describes a
random perturbation of the geodesic motion. We show that in a
spherically symmetric static background the expectation value of the
angular momentum and the energy grow exponentially fast. The model
is rather unphysical without a friction. Following our earlier paper
\cite{haba} we discuss the friction terms leading to the J\"uttner
and quantum distributions (diffusions with J\"uttner equilibrium
distribution are also discussed in \cite{chev2}\cite{bai2}). We show
that such a friction ensures finite expectation values of energy and
angular momentum.

The diffusion process describes an evolution of the particle in a
gas around a star (or a black hole ) before achieving an
equilibrium with the surrounding medium (this may be the gas of
Hawking radiation \cite{hara} from the black hole). In
ref.\cite{habampa} we studied photon diffusion in an electron gas.
Here, we discuss a soluble example of a particle diffusion in de
Sitter universe. We obtain a surprising result that the diffusion
of momentum is described by the same equation as the relativistic
diffusion on the Minkowski mass-shell with a friction leading to
the J\"uttner distribution. The corresponding temperature is
proportional to the radius of de Sitter space.  The relativistic
diffusion (without friction) in the Schwarzschild metric has been
studied earlier in \cite{lejan} and in the G\"odel universe in
\cite{fran}. For reviews on the relativistic diffusion see
\cite{chev1}\cite{han}.

The paper is organized as follows. In sec.2 we discuss the geometry
of the mass-shell. In sec.3 we construct the generator of the
diffusion on the mass-shell as the Laplace-Beltrami operator on the
level surface in the cotangent bundle.  In sec.4 we restrict
ourselves to isotropic metrics. We discuss the stochastic Ito
equations in these coordinates. In sec.5 we obtain the transport
equations in the laboratory time corresponding to the diffusion
equations in the proper time. We discuss friction forces which lead
to equilibrium measures determined by basic principles of the
statistical physics. In sec.6 a time evolution of the energy and the
angular momentum is discussed. In sec.7 we take the limit of zero
mass. The diffusion in de Sitter space  is discussed in sec.8. In
two appendices  we give a derivation of some results applying
methods of the stochastic calculus.
\section{Riemannian geometry of the mass-shell}
 We are interested in random perturbations of the dynamics of relativistic
 particles of mass $m$ moving in a gravitational field. The dynamics
 on a pseudoriemanian manifold ${\cal M}$ (the geodesic motion) is described by the equations \cite{in}
 \begin{equation}
 \frac{dx^{\mu}}{d\tau}=\frac{1}{m}g^{\mu\nu}p_{\nu},
 \end{equation}
 \begin{equation}
 \frac{dp_{\rho}}{d\tau}=-\frac{1}{2m}g^{\mu\nu}_{,\rho}p_{\mu}p_{\nu}
 \end{equation}
 where $\mu=0,1,2,3$ and
\begin{equation} ds_{x}^{2}=g_{\mu\nu}(x)dx^{\mu}dx^{\nu}
\end{equation}
is the pseudoriemannian metric on ${\cal M}$. The four-momentum  $p(\tau)$ of a
relativistic particle defines the mass by the relation
\begin{equation}
p^{2}=g^{\mu\nu}p_{\mu}p_{\nu}=m^{2}c^{2}.
\end{equation} If eq.(4) is satisfied then from eq.(1) it follows that $\tau$ has
the meaning of the proper time (which is invariant under coordinate
transformations). Eq.(4) defines a submanifold in the cotangent
bundle. We take the pseudoriemanian metric on the cotangent bundle
${\cal M}\times T{\cal M}^{*}$ as the product metric  on ${\cal M}$
and $T{\cal M}^{*}$ \begin{equation}
ds^{2}=ds_{x}^{2}+ds^{2}_{p}=g_{\mu\nu}(x)dx^{\mu}dx^{\nu}+g^{\mu\nu}(x)dp_{\mu}dp_{\nu}.
\end{equation}
The metric on the submanifold (4) is inherited from the product
metric on the cotangent bundle. It can be obtained by expressing
$p_{0}$ by the spatial components of the momenta (here
$g^{\mu\nu}$ is the inverse matrix to the one defining the metric
on $T{\cal M}$; we choose the convention that the spatial terms in
$ds^{2}$ have an opposite sign)
\begin{equation}
ds_{p}^{2}=g^{\mu\nu}(x)dp_{\mu}dp_{\nu}=g^{00}dp_{0}dp_{0}+2g^{0k}dp_{0}dp_{k}-g^{jk}dp_{j}dp_{k}
=-\gamma^{jk}dp_{j}dp_{k}\end{equation} ($j,k=1,2,3$).We assume
 that $g^{00}> 0$ and $g^{j0}=0$ (static metrics can be chosen in this form). Then,
we obtain

\begin{equation}\begin{array}{l}\gamma^{jk}=g^{jk}
-\omega^{-2}g^{jr}g^{kn}p_{r}p_{n}
\end{array}\end{equation}
We have
\begin{equation}
D=\det(\gamma^{jk})=\det(g^{jk})\omega^{-2}
\end{equation}

 and \begin{equation}\begin{array}{l}\gamma_{jk}=g_{jk}
+m^{-2}c^{-2}p_{j}p_{k}.
\end{array}\end{equation}

The measure invariant under all coordinate transformations of the
submanifold (4) of $T{\cal M}^{*}$ reads \cite{klein}
\begin{equation}
p_{0}^{-1}dxd{\bf p}.
\end{equation}
\section{Diffusion on the cotangent bundle}
We define the Laplace-Beltrami operator on the cotangent bundle as
usual by means of the metric. The metric (5) is a product of the
metrics  whereas the Laplace-Beltrami operator $\triangle$ is a sum
$\triangle_{x}+\triangle _{p}$ . The part of the Laplace-Beltrami
containing derivatives over the momenta reads
\begin{equation}
\triangle_{p}=g_{\mu\nu}\frac{\partial}{\partial p_{\mu}}
 \frac{\partial}{\partial p_{\nu}}.
 \end{equation}
 In order to define the Laplace-Beltrami operator on the
 submanifold (4) of the cotangent bundle we have to restrict the operator
 (11) to this submanifold. A way to do it is to treat the condition (4)
 as a definition of the level surface in the cotangent bundle.
 In such a case the Laplace-Beltrami operator on the level surface
 is the same as the one defined by the Riemannian geometry of the
 Riemannian mass-shell discussed in sec.2.\begin{equation}
\triangle_{H}=D^{-\frac{1}{2}}\frac{\partial}{\partial
p_{j}}\gamma_{jk}D^{\frac{1}{2}}\frac{\partial}{\partial p_{k}},
\end{equation}where $ D$ is defined in eq.(8) and $\gamma_{jk}$ in eq.(7).

Explicitly,
\begin{equation}\begin{array}{l}
\triangle_{H}=
(g_{jk}+m^{-2}c^{-2}p_{j}p_{k})\frac{\partial}{\partial
p_{j}}\frac{\partial}{\partial
p_{k}}+3m^{-2}c^{-2}p_{k}\frac{\partial}{\partial p_{k}}.
\end{array}\end{equation}
The  diffusion equation is
\begin{equation}
\partial_{\tau}\phi_{\tau}={\cal G}\phi_{\tau}
\end{equation}
with the generator
\begin{equation}
{\cal G}=g^{\mu\nu}p_{\mu}\frac{\partial}{\partial
x^{\nu}}-\frac{1}{2m}g^{\mu\nu}_{,j}p_{\mu}p_{\nu}\frac{\partial}{\partial
p_{j}}+\frac{\kappa^{2}m^{2}c^{2}}{2}\triangle_{H}.
\end{equation}
 The operator (15) generates a diffusive dynamics
which is a perturbation of the geodesic motion (1)-(2) by a
diffusion in the momentum space. A coordinate free frame bundle
definition of the diffusion (14) has been obtained earlier in
\cite{lejan}(Lemma 3.1).

 We still transform eq.(15) in order to
express the diffusion on the Riemannian mass-shell as a
perturbation of a diffusion on the Minkowski mass-shell (we apply
this transformation in sec.8). For this purpose let us introduce
the tetrads $f_{j}^{a}$ by the equation
\begin{equation}
g_{jl}=f^{a}_{j}f^{a}_{l}
\end{equation}($a=1,2,3$).
We define the inverse $f_{a}^{j}$
\begin{equation}
f_{a}^{j}f^{b}_{j}=\delta_{a}^{b}.
\end{equation}
We change coordinates on the cotangent bundle
\begin{equation}
p_{j}=f^{a}_{j}p_{a}^{\prime}
\end{equation} while  $x^{\nu}=x^{\prime \nu}$.
 The coordinate vector fields take the form
\begin{equation} \frac{\partial}{\partial p_{j}}=
f_{a}^{j}\frac{\partial}{\partial p_{a}^{\prime}},\end{equation}

\begin{equation}
\frac{\partial}{\partial x^{\nu}}=\frac{\partial}{\partial
x^{\nu\prime}}+\omega_{\nu
a}^{b}p_{b}^{\prime}\frac{\partial}{\partial p_{a}^{\prime}},
\end{equation}
where
\begin{equation}\omega_{\nu
a}^{b}= \frac{\partial f_{a}^{l}}{\partial
x^{\prime\nu}}f_{l}^{b}.
\end{equation}
The vector field  of eq.(20) can be considered as a horizontal
lift of the coordinate vector field $\partial_{\nu}$ from ${\cal
M}$ to ${\cal M}\times T{\cal M}^{*}$.
\section{Relativistic diffusion in isotropic coordinates}
We shall consider highly  symmetric manifolds. Such manifolds admit
special coordinate systems. We restrict ourselves to the isotropic
spherically symmetric metric\cite{in} from now on. Then,
\begin{equation}
ds^{2}_{x}=A^{-2}(\vert{\bf x}\vert)c^{2}dt^{2}-B^{-2}(\vert{\bf
x}\vert)d{\bf x}^{2}.
\end{equation}In these coordinates
\begin{equation}\begin{array}{l}
\triangle_{H}=(B^{-2}\delta_{jk}+m^{-2}c^{-2}p_{j}p_{k})\frac{\partial}{\partial
p_{j}}\frac{\partial}{\partial
p_{k}}+3m^{-2}c^{-2}p_{k}\frac{\partial}{\partial p_{k}}.
\end{array}\end{equation}
The diffusion equation in the isotropic coordinates reads
\begin{equation}\begin{array}{l}
\partial_{\tau}\phi=\frac{1}{m}A\omega\frac{\partial}{\partial
x^{0}}\phi-\frac{1}{m}B^{2}p_{j}\frac{\partial}{\partial
x^{j}}\phi +\frac{1}{2m}\frac{\partial B^{2}}{\partial x^{j}}{\bf
p}^{2}\frac{\partial \phi}{\partial
p_{j}}\cr-\frac{1}{2m}\frac{\partial \ln A}{\partial
x^{j}}\omega^{2}\frac{\partial \phi}{\partial p_{j}}
+\frac{\kappa^{2}m^{2}c^{2}}{2}\triangle_{H}\phi
\end{array}\end{equation}
where in the generator (15) we expressed $p_{0}$ as
\begin{equation}
p_{0}=A^{-1}(B^{2}{\bf p}^{2}+m^{2}c^{2})^{\frac{1}{2}}\equiv
A^{-1}\omega.
\end{equation}
We can express the solution of the diffusion equation (24) as an
expectation value $E[..]$ over the sample paths of  a diffusion
process $(x_{\tau},{\bf p}_{\tau})$ starting from $(x,{\bf p})$
\cite{ikeda}
\begin{equation}
\phi_{\tau}(x,{\bf p})=E[\phi(x_{\tau},{\bf p}_{\tau})].
\end{equation}
The stochastic process can be defined as a solution of a stochastic
equation. In order to write down the stochastic equations we must
calculate the square root of the matrix $(\gamma_{jl})$, i.e., to
find a tetrad $e$
\begin{equation}
e_{j}^{ a}e_{l}^{a}=\gamma_{jl}.
\end{equation}
We obtain
\begin{equation}
e_{j}^{a}=B^{-1}(\delta_{j}^{a}+\frac{1}{mc}(\omega-mc){\bf
p}^{-2}p_{j}p_{a}).
\end{equation}
Then, Ito stochastic differential equations \cite{ikeda}
corresponding to the representation (26) of the solution of the
diffusion equation (24)  read (see Appendix A;stochastic equations
for the relativistic diffusion have been obtained also in the
frame bundle formalism of ref.\cite{lejan},Theorem 3.2)
\begin{equation}
 \frac{dx^{0}}{d\tau}=\frac{1}{m}A\omega
 \end{equation}
\begin{equation}
\frac{d{\bf x}}{d\tau}=-\frac{1}{m}B^{2}{\bf p}
\end{equation}
\begin{equation}\begin{array}{l} dp_{j}=\frac{1}{
m}B\partial_{j}^{x}B{\bf
p}^{2}d\tau-\frac{1}{m}\omega^{2}\partial_{j}^{x}\ln
Ad\tau+\frac{3\kappa^{2}}{2}p_{j}d\tau \cr +mc\kappa B^{-1}db_{j}+
\kappa B^{-1}(\omega-mc){\bf p}^{-2}p_{j}{\bf p}d{\bf b}.
\end{array}\end{equation}
Here, ${\bf b}(s)$ denotes the Brownian motion defined as the
Gaussian process with values in $R^{3}$ and the covariance
\begin{displaymath}
E[b_{a}(s)b_{c}(\tau)]=\delta_{ac}min(s,\tau)
\end{displaymath}

\section{Invariant measure and kinetic equations} We define an
evolution of the measure $d\sigma=dxd{\bf p}\Phi$ by the equality
(see
\cite{haba})\begin{equation}\langle\phi_{\tau}\rangle_{\sigma}=\int
dxd{\bf p}\Phi \phi_{\tau}=\int dxd{\bf p}\Phi_{\tau} \phi.
\end{equation}
Then, \begin{equation} \partial_{\tau}\Phi_{\tau}={\cal
G}^{*}\Phi_{\tau}\end{equation}where ${\cal G}^{*}$ is the adjoint
of ${\cal G}$ in $L^{2}(dxd{\bf p})$. We say that the measure
$\sigma $ is invariant if
\begin{equation} \langle
\phi_{\tau}\rangle_{\sigma}=\langle \phi\rangle_{\sigma}
\end{equation}is independent of $\tau$. In such a case
\begin{equation}
{\cal G}^{*}\Phi=0.\end{equation} Eq.(35) is a transport equation
in the laboratory time $x^{0}$. A normalizable invariant measure
needs a dissipation. We add such a dissipation term $K_{j}d\tau$
to the stochastic equation (31). This is equivalent to adding the
friction

\begin{equation}
K=K_{j}\frac{\partial}{\partial p_{j}} \end{equation} to the
diffusion equation (24).

 In isotropic coordinates eq.(35) reads
\begin{equation}\begin{array}{l}
-\frac{1}{m}A^{2}p_{0}\frac{\partial}{\partial
x^{0}}\Phi+\frac{1}{m}\frac{\partial}{\partial x^{j}}B^{2}p_{j}\Phi
-\frac{1}{2m}\frac{\partial B^{2}}{\partial x^{j}}\frac{\partial
}{\partial p_{j}}{\bf p}^{2}\Phi+\frac{1}{2m}\frac{\partial
A^{2}}{\partial x^{j}}\frac{\partial }{\partial
p_{j}}p_{0}^{2}\Phi\cr +
\frac{\kappa^{2}m^{2}c^{2}}{2}\frac{\partial}{\partial
p_{j}}\frac{\partial}{\partial
p_{k}}(B^{-2}\delta_{jk}+m^{-2}c^{-2}p_{j}p_{k})\Phi-\frac{3\kappa^{2}}{2}\frac{\partial}{\partial
p_{k}}p_{k}\Phi-\frac{\partial}{\partial p_{k}}K_{k}\Phi=0.
\end{array}
\end{equation}
In the search of the equilibrium distribution $\Phi_{E}$ we demand
that the dynamic part and the diffusion part of the  equation (37)
cancel separately. So, for the dynamic part
\begin{equation}\begin{array}{l}
-\frac{1}{m}A^{2}p_{0}\frac{\partial}{\partial
x^{0}}\Phi_{E}+\frac{1}{m}\frac{\partial}{\partial
x^{j}}B^{2}p_{j}\Phi_{E} -\frac{1}{2m}\frac{\partial
B^{2}}{\partial x^{j}}\frac{\partial }{\partial p_{j}}{\bf
p}^{2}\Phi_{E}+\frac{1}{2m}\frac{\partial A^{2}}{\partial
x^{j}}\frac{\partial }{\partial p_{j}}p_{0}^{2}\Phi_{E}=0.
\end{array}
\end{equation}Then, from the diffusion part \begin{equation}\begin{array}{l}
K_{k}=\kappa^{2}m^{2}c^{2}\Phi_{E}^{-1}\Big(\frac{1}{2}\frac{\partial}{\partial
p_{j}}(B^{-2}\delta_{jk}+m^{-2}c^{-2}p_{j}p_{k})-\frac{3}{2}(mc)^{-2}p_{k}\Big)\Phi_{E}.\end{array}
\end{equation}Any function of deterministic constants
of motion is a solution of the dynamic equation (38) (static
metric with  a spherical symmetry has the angular momentum and the
energy as constants of motion). Let us consider the solution as a
function of $p_{0}$ (25) in the form ( J\"uttner \cite{jut}
 and quantum statistical distributions are of this form)
\begin{equation}
\Phi_{E}=A\omega^{-1}\exp(f(\beta cA^{-1}\omega)).
\end{equation}
 From eq.(39) we obtain
\begin{equation}
K_{k}=\frac{\kappa^{2}}{2}p_{k}\beta cA^{-1}\omega
f^{\prime}(\beta c A^{-1}\omega).
\end{equation}(diffusions with J\"uttner equilibrium distribution have
been discussed also in \cite{chev2}\cite{bai2}). Explicitly, with
the friction (39) the diffusion equation is
\begin{equation}\begin{array}{l}
\partial_{\tau}\phi=\frac{1}{m}A\omega\frac{\partial}{\partial
x^{0}}\phi-\frac{1}{m}B^{2}p_{j}\frac{\partial}{\partial
x^{j}}\phi +\frac{1}{2m}\frac{\partial B^{2}}{\partial x^{j}}{\bf
p}^{2}\frac{\partial \phi}{\partial
p_{j}}-\frac{1}{2m}A^{-2}\frac{\partial A^{2}}{\partial
x^{j}}\omega^{2}\frac{\partial \phi}{\partial p_{j}} \cr
+\frac{\kappa^{2}m^{2}c^{2}}{2}(B^{-2}\delta_{jk}+m^{-2}c^{-2}p_{j}p_{k})\frac{\partial}{\partial
p_{j}}\frac{\partial}{\partial
p_{k}}\phi+\kappa^{2}(\frac{3}{2}+\frac{1}{2}\omega c\beta
A^{-1}f^{\prime}(\beta c\omega
A^{-1}))p_{k}\frac{\partial}{\partial p_{k}}\phi.
\end{array}\end{equation}
Then, the stochastic equation for the process ${\bf p}_{\tau}$ (26)
solving the diffusion equation (42) reads (see Appendix A)
\begin{equation}\begin{array}{l}
dp_{j}=\frac{1}{m}B\partial_{j}^{x}B{\bf
p}^{2}d\tau-\frac{1}{m}\omega^{2}\partial_{j}^{x}\ln Ad\tau\cr
+\frac{p_{j}\kappa^{2}}{2}\omega c\beta
A^{-1}f^{\prime}d\tau+\frac{3\kappa^{2}}{2}p_{j}d\tau +mc\kappa
B^{-1}db_{j}+ \kappa B^{-1}(\omega-mc){\bf p}^{-2}p_{j}{\bf
p}d{\bf b}.
\end{array}\end{equation}
The equilibrium distributions (40) (J\"uttner or quantum) are not
invariant under Lorentz transformations even in the flat
(Minkowski) case. The notion of the equilibrium depends on the
Lorentz frame where the particle equilibrates.  We make this frame
dependence explicit in \cite{habajmp}.

\section{Evolution of the energy and the angular momentum}
In the isotropic coordinates we define the angular momentum
\begin{equation}
{\bf L}={\bf x}\times {\bf p}.
\end{equation}
If $A(\vert{\bf x}\vert)$ and $B(\vert{\bf x}\vert)$  depend only
on $\vert{\bf x}\vert$ then the metric is spherically symmetric.
In such a case when $\kappa=0$ then the angular momentum is a
constant of motion. First, we consider the proper time evolution
of the angular momentum in a stochastic model without a friction.
From eq.(31) we obtain a stochastic equation for the time
evolution of the angular momentum when $\kappa\neq 0$ (see
Appendix B)
\begin{equation}
d{\bf L}_{\tau}=\frac{3\kappa^{2}}{2}{\bf L}_{\tau}d\tau +\kappa
B^{-1}(\omega-mc){\bf p}^{-2}{\bf L}_{\tau}({\bf p}d{\bf
b})+\kappa mc B^{-1}{\bf x}\times d{\bf b}.
\end{equation}
In the spherical coordinates $(r,\theta,\phi)$ eq.(45) can be
expressed in the form
\begin{equation} dp_{\phi}=
\frac{3\kappa^{2}}{2}p_{\phi}d\tau+\kappa e^{a}_{\phi}db_{a}.
\end{equation}
We write eqs.(45)-(46) in an integral form and apply the basic
property of the Ito integral \cite{ikeda}\begin{equation}E[ \int
Fdb]=0.
\end{equation}
Then, it follows from eq.(45) that
\begin{equation}\exp(-\frac{3\kappa^{2}}{2}\tau)E[{\bf
L}_{\tau}]=const.
\end{equation}
Eq.(48) means that the angular momentum grows exponentially in
time.

In a static metric when $\kappa=0$    the energy
$p_{0}=A^{-1}\omega$ is also a constant of motion. We calculate
the change of energy during the proper time diffusion (31) in a
model without friction.
  Differentiating eq.(25) we obtain (see Appendix B)
 \begin{equation}\begin{array}{l}
 dp_{0}=\frac{3}{2}\kappa^{2}p_{0}d\tau+\kappa A^{-1}B{\bf p}d{\bf
 b}.
\end{array} \end{equation}
 It follows from eqs.(49) and (47) that
\begin{equation}\begin{array}{l}
 {\cal E}(\tau)=\frac{3}{2}\kappa^{2}\int_{0}^{\tau}{\cal E}(s)ds
 \end{array}\end{equation}
where
\begin{displaymath}
{\cal E}(s)=E[p_{0}(s)]. \end{displaymath} Eq.(50) has the
solution
\begin{equation}
{\cal E}(\tau)={\cal E}(0)\exp(\frac{3}{2}\kappa^{2}\tau)
\end{equation}
We  calculate a change of the angular momentum (in a spherically
symmetric metric) resulting from the diffusion (43) with friction
(Appendix B)
\begin{equation}
d{\bf L}_{\tau}=\frac{\kappa^{2}}{2}\omega c\beta
A^{-1}f^{\prime}{\bf L}d\tau+\frac{3\kappa^{2}}{2}{\bf
L}_{\tau}d\tau +\kappa B^{-1}(\omega-mc){\bf p}^{-2}{\bf
L}_{\tau}({\bf p}d{\bf b})+\kappa mc B^{-1}{\bf x}\times d{\bf b}.
\end{equation}
$f^{\prime}$ is negative for J\"uttner as well as for quantum
equilibrium distributions so that the angular momentum does not
increase as it did in eq.(48). In the spherical coordinates we
have
\begin{equation}
dp_{\phi}=\frac{p_{\phi}\kappa^{2}}{2}\omega c\beta
A^{-1}f^{\prime}d\tau+\frac{3\kappa^{2}}{2}p_{\phi}d\tau
+e^{a}_{\phi}db_{a} .\end{equation} $\omega$ is growing linearly
with $p_{\phi}$ for large $p_{\phi}$. For J\"uttner distribution
$f^{\prime}=-1$ and for large $p_{\phi}$ eq.(53) is of the form
\begin{equation}
dp_{\phi}=\alpha p_{\phi} d\tau-\delta
p_{\phi}^{2}d\tau+e^{a}_{\phi}db_{a}.
\end{equation}
The stochastic equation for the energy is also of the form (54). We
have from eq.(43) (see Appendix B)
 \begin{equation}\begin{array}{l}
 dp_{0}=\frac{3}{2}\kappa^{2}p_{0}d\tau+
\frac{\kappa^{2}}{2}c\beta
p_{0}^{2}f^{\prime}d\tau-c\beta\frac{\kappa^{2}m^{2}c^{2}}{2}A^{-2}f^{\prime}d\tau
+\kappa A^{-1}B{\bf p}d{\bf
 b}.\end{array}
\end{equation}
We are going to prove that the expectation value of the energy is
bounded in time. For this purpose  the Lyapunov methods
\cite{hasm}\cite{las} are needed. We must find a non-negative increasing to infinity
Lyapunov function $V$ such that for  constants $a\geq 0$ and  $K>0$
\begin{equation}
{\cal G}V\leq -K V+a.
\end{equation}
The lhs of eq.(56) is a time derivative of $E[V({\bf p}_{\tau})]$.
The integral form
 of the inequality (56) reads \cite{las}
 \begin{equation}
E[V({\bf p}_{\tau})]\leq\exp(-K\tau)V({\bf
p})+\frac{a}{K}(1-\exp(-K\tau)).
\end{equation}
We choose $V({\bf p})=p_{0}=A^{-1}\omega$. Then,
 by direct calculations
 \begin{equation}\begin{array}{l}
 {\cal G}p_{0}=\frac{3}{2}\kappa^{2}p_{0}+
\frac{\kappa^{2}}{2}c\beta
p_{0}^{2}f^{\prime}-c\beta\frac{\kappa^{2}m^{2}c^{2}}{2}A^{-2}f^{\prime}.
 \end{array}
\end{equation}
Eq.(58) is also a consequence of eqs.(26),(47) and (55) as
\begin{displaymath}
dE[p_{0}]=E[dp_{0}]={\cal G}p_{0}d\tau
\end{displaymath}
If $f^{\prime}<0$ (true for J\"uttner as well as quantum
equilibrium distributions) then from eq.(58) the inequality (56)
results. As a consequence of eq.(57) the energy is bounded in
time. The proof assumes  that the stochastic equation (43) for
${\bf p}$ has solutions for arbitrary time. If $ A$ and $B^{-1}$
are bounded functions then an extension of the solution to
arbitrarily large time can easily be shown by means of the
Lyapunov function method (choose $V({\bf p})={\bf p}^{2}$ as the
Lyapunov function).  We could also apply the Lyapunov argument to
the stochastic equation for the angular momentum (53) assuming
that $A$ and $B^{-1}$ are bounded ( with $p_{\phi}^{2}$ as the
Lyapunov function).  The problem with singular $A $ and $B^{-1}$
(the case of the Schwarzschild solution) is more complicated. The
diffusion can terminate in the singularity at finite time. The
properly behaving diffusion can be constructed on the Kruskal
extension of the Schwarzschild solution \cite{lejan}.
 Summarizing, without the friction the expectation value of the
energy and the angular momentum  grow exponentially fast whereas with the friction, leading
to the J\"uttner or Bose-Einstein equilibrium distribution, the
energy and the angular momentum are bounded in time.

We perform some explicit (although approximate) calculations of
the expectation value of the energy for
 the J\"uttner equilibrium distribution ($f^{\prime}=-1$).
 From eq.(55) we
obtain an equation
\begin{equation}\begin{array}{l}
 {\cal E}(\tau)=\frac{3}{2}\kappa^{2}\int_{0}^{\tau}ds{\cal E}(s)
-\frac{\kappa^{2}}{2}c\beta \int_{0}^{\tau}dsE[p_{0}(s)^{2}] \cr
+\frac{1}{2}c\beta\kappa^{2}m^{2}c^{2}\int_{0}^{\tau}dsE[A^{-2}].
\end{array}
\end{equation}
Eq.(59) is not a closed equation for the expectation value of the
energy because $p_{0}$ is inside the expectation values on the rhs
of eq.(59). We need a perturbative method. A linearization of
eqs.(43) or (55) is not a proper tool because the $p_{0}^{2}$ term
is crucial for the equilibration. We can apply an expansion which
assumes small $\kappa$ and  finite $\kappa^{2}c\beta$ and
$\kappa^{2}m^{2}c^{2}$. In such an expansion the stochastic force
in eqs.(43) and (55) can be neglected in the lowest order (
another method often applied in stochastic equations is based on
time averaging which eliminates stochastic forces at the lowest
order). Then,
\begin{equation}
E[p_{0}^{2}]\simeq E[p_{0}]^{2} \end{equation}  ( in general
$E[p_{0}^{2}]\geq E[p_{0}]^{2}$ ). Eq.(59) with the approximation
(60) is equivalent to the differential equation
 \begin{equation}\begin{array}{l}
 \frac{d}{d\tau}{\cal E}(\tau)=\frac{3}{2}\kappa^{2}{\cal E}(\tau)
-\frac{\kappa^{2}}{2}c\beta {\cal
E}(\tau)^{2}+\frac{1}{2}c\beta\kappa^{2}m^{2}c^{2}A^{-2}.
 \end{array}
\end{equation}There is still $p_{0}$ in the argument of $A^{-2}$.
Applying the same reasoning as in eq.(60)(expectation value of a
function is approximately a function of the expectation value) we
would obtain a closed integro-differential equation (61) for ${\cal
E}(\tau)$. Such an equation could be approached by iterative
methods. We make a simplifying assumption that $A^{-2}$ is a slowly
varying function of its argument
\begin{equation}
E[A^{-2}({\bf x}_{\tau})]\simeq A^{-2}({\bf x})
\end{equation}
where the rhs does not depend on time.
 The solution of eq.(61) with the approximation (62) is
\begin{equation}\begin{array}{l}
{\cal E}(\tau)=\Big(({\cal
E}(0)-\epsilon_{-})\epsilon_{+}\exp(\tau\kappa^{2}(\epsilon_{+}-\epsilon_{-}))
-\epsilon_{-}({\cal E}(0)-\epsilon_{+})\Big)\cr\Big(({\cal
E}(0)-\epsilon_{-})\exp(\tau\kappa^{2}(\epsilon_{+}-\epsilon_{-}))
-{\cal E}(0)+\epsilon_{+}) \Big)^{-1}\end{array}
\end{equation}
where
\begin{equation}
\epsilon_{\pm}=\frac{3}{2c\beta}\pm\Big(m^{2}c^{2}A^{-2}+\frac{9}{4c^{2}\beta^{2}}\Big)^{\frac{1}{2}}.
\end{equation}
In the limit $\tau\rightarrow\infty$ we obtain
\begin{equation}
{\cal E}(\infty)=
\epsilon_{+}=\frac{3}{2c\beta}+\Big(m^{2}c^{2}A^{-2}+\frac{9}{4c^{2}\beta^{2}}\Big)^{\frac{1}{2}}
\end{equation}
In the  limit of  large $\beta$ and for the Schwarzschild solution
\cite{in} ($G$ is the Newton constant, $M$ is the mass of the source
of gravity)
\begin{equation}
{\cal E}(\infty)\simeq\frac{3}{2c\beta}+mcA^{-1}\simeq\frac{3}{2c\beta}+mc-\frac{GMm}{rc}.
\end{equation}
This is the classical non-relativistic equipartition of energy in a
gravitational potential $\frac{M}{r}$. In the limit of zero mass
\begin{displaymath}{\cal E}(\infty)=\frac{3}{c\beta}=\Big(\int d{\bf p}\exp(-c\beta\vert{\bf p})\vert\Big)^{-1}\int
d{\bf p}\vert {\bf p}\vert\exp(-c\beta\vert{\bf p}\vert )
   \end{displaymath}
we obtain the relativistic equipartition of energy
$\epsilon=c{\cal E}(\infty)$ of massless particles (the result is
correct and could be derived rigorously on the basis of an exact
solution of eq.(80) and a calculation of its expectation values in
the next section).

The approach to the equilibrium is exponential with the speed
$\kappa^{2}(\epsilon_{+}-\epsilon_{-})$. An exponential decay to
the equilibrium of relativistic diffusions  could be proved by
means of general methods of stochastic differential equations as
discussed in \cite{bai2}, sec.3.2. Another method could be based
on our results in \cite{haba}. We have shown in \cite{haba}
(sec.9) that the time evolution of the relativistic diffusion can
be described as the time evolution in imaginary time quantum
mechanics. The approach to the equilibrium is equivalent to the
approach to the ground state (this is shown explicitly in the
model of \cite{habampa}) with the speed equal to the eigenvalue of
the first excited state of the Hamiltonian. There are
well-developed methods in quantum mechanics for a study of this
eigenvalue.

\section{The limit $m\rightarrow 0$}
There is a substantial simplification of stochastic equations in
the limit $m\rightarrow 0$. We have studied this diffusion on the
Minkowski space-time in \cite{habampa}. We discuss a
representation of the Poincare group for diffusing massless
particles in \cite{habajpa}. It is shown that  the helicity does
not mix with the diffusion. Hence, our equations without a spin
can describe diffusion of photons as well as the ultrarelativistic
behaviour of massive particles with a spin (so it can be useful in
a description of heavy ion collisions \cite{heavy}). The limit
$m\rightarrow 0$ seems singular when applied to the generator
(13). However, when we consider the time evolution $\exp(\tau
m^{2}\triangle_{H})$ then the limit $m\rightarrow 0$ exists (it
corresponds to a rescaling of an affine time parameter which in
the massless case  does not have the meaning of the proper time
anyhow). The diffusion generator in the limit $m\rightarrow 0$
reads
\begin{equation}\begin{array}{l}
\triangle_{H}= p_{j}p_{k}\frac{\partial}{\partial
p_{j}}\frac{\partial}{\partial
p_{k}}+3p_{k}\frac{\partial}{\partial p_{k}}.
\end{array}\end{equation}  The differential
operator (67) defined on the manifold ${\cal M}$ does not depend on
the metric. It is degenerate as an elliptic operator because the
quadratic form $(p_{j}a_{j})^{2}$ is degenerate. As a consequence of
this degeneration the diffusion in the configuration space is
trivial. The time evolution of a massless particle on the Minkowski
space describes a deterministic motion with a velocity of light on a
straight line (as discussed in more detail in \cite{habajmp}). For
this reason it has been rejected by Dudley \cite{dudley} (sec.11).
However, in the momentum space it describes a diffusion of the
energy. We have shown in \cite{habampa} that the diffusion of
massless particles is just a linearization of the Kompaneetz
diffusion well-known in astrophysics \cite{rybicki}. On a curved
manifold the motion of a massless particle in configuration space is
non-trivial but still deterministic.The stochastic Ito differential
equations (29)-(31) take the simple form
  \begin{equation}
 \frac{dx^{0}}{d\tau}=AB\vert{\bf p}\vert,
 \end{equation}
\begin{equation}
\frac{d{\bf x}}{d\tau}=-B^{2}{\bf p},
\end{equation}
\begin{equation}\begin{array}{l} dp_{j}=B\partial_{j}^{x}B{\bf
p}^{2}d\tau-\omega^{2}\partial_{j}^{x}\ln
Ad\tau+\frac{3\kappa^{2}}{2}p_{j}d\tau+ \kappa p_{j}db.
\end{array}\end{equation}
There is only one Brownian motion $b$ for all $j$ components of
$p_{j}$ ( a consequence of the degenerate generator (67)).

 The transport equation reads
\begin{equation}\begin{array}{l}
-A^{2}p_{0}\frac{\partial}{\partial
x^{0}}\Phi+\frac{\partial}{\partial x^{j}}B^{2}p_{j}\Phi
-\frac{\partial B^{2}}{\partial x^{j}}\frac{\partial }{\partial
p_{j}}{\bf p}^{2}\Phi+\frac{\partial A^{2}}{\partial
x^{j}}\frac{\partial }{\partial p_{j}}p_{0}^{2}\Phi\cr +
\frac{\kappa^{2}}{2}\frac{\partial}{\partial
p_{j}}\frac{\partial}{\partial
p_{k}}p_{j}p_{k}\Phi-\frac{3\kappa^{2}}{2}\frac{\partial}{\partial
p_{k}}p_{k}\Phi-\frac{\partial}{\partial p_{k}}K_{k}\Phi=0.
\end{array}
\end{equation} where the friction $K_{k}$ is determined by the equilibrium
distribution $\Phi_{E}$
\begin{equation}\begin{array}{l}
K_{k}=\kappa^{2}\Phi_{E}^{-1}\Big(\frac{1}{2}\frac{\partial}{\partial
p_{j}}p_{j}p_{k}-\frac{3}{2}p_{k}\Big)\Phi_{E}.\end{array}
\end{equation}
We  discuss two examples of the equilibrium measures
\begin{equation}
\Phi_{L}=p_{0}^{-1}\exp(-\beta_{\phi}p_{\phi})
\end{equation}
and\begin{equation} \Phi_{E}=p_{0}^{-1}\exp(-\beta p_{0}).
\end{equation}
In the first case
\begin{equation}
K_{k}=-\frac{\kappa^{2}}{2}\beta_{\phi}p_{k}p_{\phi}.
\end{equation}
Hence, the stochastic equation for the angular momentum reads
\begin{equation}
dp_{\phi}=\frac{3\kappa^{2}}{2}p_{\phi}d\tau-\frac{\kappa^{2}}{2}\beta_{\phi}p_{\phi}^{2}d\tau
+\kappa p_{\phi}db.
\end{equation}
Let us write
\begin{equation}
p_{\phi}=\exp(u).
\end{equation}
The solution of eq.(76) is
\begin{equation}
u_{\tau}=u+\kappa^{2}\tau+\kappa
b_{\tau}-\ln\Big(1+\frac{\kappa^{2}}{2}\beta_{\phi}\exp(u)\int_{0}^{\tau}
ds\exp(\kappa^{2}s+\kappa b_{s})\Big).
\end{equation}
In the  case (74) we obtain
\begin{equation}
K_{k}=-\frac{\kappa^{2}}{2}\beta p_{k}p_{0}.
\end{equation}
Hence, the equation for $p_{0}$ reads
\begin{equation}
dp_{0}=\frac{3\kappa^{2}}{2}p_{0}d\tau-\frac{\kappa^{2}}{2}\beta
p_{0}^{2}d\tau +\kappa p_{0}db.
\end{equation}
It has the same solution as eq.(76) ( just replace $\beta_{\phi}$
by $\beta$ in eq.(78)). The correlation functions of polynomials
of $p_{0}(s)^{-1}$ and $p_{\phi}(s)^{-1} $ can be calculated
explicitly. The expectation values have a limit $\tau \rightarrow
\infty$ expressed as moments of the invariant measures (73)-(74).
The stochastic equations for $p_{\phi}$ and $p_{0}$ do not depend
on the metric. Their solutions and expectation values are
discussed in \cite{habajmp}.\section{ A particle diffusing on de
Sitter space} In this section we discuss a relativistic particle
diffusing on a background metric of the cosmological type
\begin{equation}
ds^{2}=(dx^{0})^{2}-B^{-2}(d{\bf x})^{2}.
\end{equation}It is useful to apply  the transformations (19)-(20) in order
to reduce the stochastic equations to a perturbation of the
diffusion on the Minkowski mass-shell. For simplicity of the
formulae we restrict ourselves to $B$ which depends only on time.
Then, the diffusion equation (13)-(14) after the transformation
(19)-(20) has the generator
\begin{equation}\begin{array}{l}
{\cal G}=\frac{1}{2}\kappa^{2}m^{2}c^{2}\triangle_{H}^{ Min}+
\frac{1}{m}p_{0}\frac{\partial}{\partial x^{
0}}-\frac{1}{m}Bp_{k}\frac{\partial}{\partial
p_{k}}+\frac{1}{m}p_{0}\frac{\partial B}{\partial x^{
0}}B^{-1}p_{k}\frac{\partial}{\partial p_{k}}
\end{array}\end{equation}
  where
\begin{equation}\begin{array}{l}
\triangle_{H}^{Min}=(\delta_{jk}+m^{-2}c^{-2}p_{j}p_{k})\frac{\partial}{\partial
p_{j}}\frac{\partial}{\partial
p_{k}}+3m^{-2}c^{-2}p_{k}\frac{\partial}{\partial p_{k}}
\end{array}\end{equation}(expressed in eqs.(19)-(20) by primed momenta) is the generator of the diffusion on the
Minkowski mass-shell. In eq.(82)
\begin{equation}
p_{0}=\sqrt{{\bf p}^{ 2}+m^{2}c^{2}}\equiv \omega_{1}.
\end{equation}
where $\omega_{1}$ denotes the energy on the Minkowski space-time
(with the spatial metric tensor equal to 1).

In the generator of the diffusion (82)  the diffusion part remains
the same as on the Minkowski mass-shell. The motion on the de
Sitter space is described by the drift. So, the classical geodesic
equations, when expressed in the coordinates (19)-(20), are
perturbed by the relativistic Brownian motion defined on the
mass-shell $p_{0}^{2}-{\bf p}^{2}=m^{2}c^{2}$.

The stochastic equations for the diffusion  with the generator
(82) read
\begin{equation}
\frac{dx^{0}}{d\tau}=\frac{1}{m}\omega_{1},
\end{equation}
\begin{equation}\frac{d{\bf x}}{d\tau}=-\frac{B^{2}}{m}{\bf p}
\end{equation}
and
\begin{equation}\begin{array}{l}
dp_{j}=\frac{1}{m}\omega_{1}\frac{\partial \ln B}{\partial
x^{0}}p_{j}d\tau  +\frac{3\kappa^{2}}{2}p_{j}d\tau +mc\kappa
db_{j}+ \kappa (\omega_{1}-mc){\bf p}^{-2}p_{j}{\bf p}d{\bf b}
\end{array}\end{equation}
In general, it is difficult to derive an explicit  solution of such
diffusion equations. There is one remarkable exception if (this is
the metric for a causally connected part of de Sitter space
sometimes called the Bondi-Hoyle universe \cite{in})

\begin{equation}
ds^{2}=(dx^{0})^{2}-\exp(\frac{2}{cR}x^{0})d{\bf x}^{2}
\end{equation}
($R$ is the radius of the pseudosphere in the embedding of de
Sitter space in a fivedimensional Minkowski space).

Then, the drift in eq.(87) corresponding to the metric (88) is
\begin{equation}
K_{j}=-\frac{1}{mcR}\omega_{1} p_{j}
\end{equation}
exactly the same  as the one for the relativistic model on the
Minkowski mass-shell (eq.(42) $B=A=1$) with the friction leading
to the J\"uttner equilibrium distribution ($f^{\prime}=-1$) and
the inverse temperature
\begin{equation}
\kappa^{2}\beta=\frac{2}{mc^{2}R}.
\end{equation}
As $\frac{1}{R }=\sqrt{\frac{1}{3}\Lambda} $, we obtain a relation
between the temperature and the cosmological constant $\Lambda$.
Eq.(37) for the equilibrium distribution has now the solution
\begin{equation}
\Phi_{E}({\bf
p})=\omega_{1}^{-1}\exp\Big(-\omega_{1}\frac{2}{m\kappa^{2}c^{2}}\sqrt{\frac{1}{3}\Lambda}\Big)
\end{equation}
The model of a massless particle diffusing in de Sitter space is
explicitly soluble. Eqs.(85)-(87) read
\begin{equation}
\frac{dx^{0}}{d\tau}=\vert {\bf p}\vert,
\end{equation}
\begin{equation}\frac{d{\bf x}}{d\tau}=-\exp(-\frac{2x^{0}}{cR}){\bf p}
\end{equation}
and
\begin{equation}\begin{array}{l}
dp_{j}=-\frac{1}{cR}\vert{\bf p}\vert p_{j}d\tau
+\frac{3\kappa^{2}}{2}p_{j}d\tau +\kappa p_{j}db.
\end{array}\end{equation}
From eq.(94) it follows that ${\bf p}_{\tau}=\vert{\bf
p}_{\tau}\vert {\bf n}$, where ${\bf n}$ is a time independent
unit vector. The equation for $\vert {\bf p}\vert$ is the same as
eq.(76) with the solution
\begin{equation}
\ln \vert {\bf p}_{\tau}\vert=\ln \vert{\bf
p}\vert+\kappa^{2}\tau+\kappa
b_{\tau}-\ln\Big(1+\frac{1}{Rc^{2}}\vert {\bf
p}\vert\int_{0}^{\tau} ds\exp(\kappa^{2}s+\kappa b_{s})\Big).
\end{equation}
 We can
calculate expectation values of functions of ${\bf p}_{\tau}$
using the formulae of Yor \cite{yor}  for the probability
distribution of exponentials of the Brownian motion and the
integrals of exponentials (we have calculated some expectation
values in \cite{habajmp}).  The correlation functions of $\vert
{\bf p}_{\tau}\vert^{-1}$ can be expressed by elementary
functions. The equilibrium measure reads
\begin{displaymath}
\Phi_{E}({\bf p})=\vert {\bf
p}\vert^{-1}\exp(-\frac{2}{R\kappa^{2}c}\vert {\bf p}\vert).
\end{displaymath}
It is rather surprising that the momentum evolution in de Sitter
space tends to an equilibrium as if a certain friction was
involved in a geodesic diffusion on de Sitter space.
\section{Summary} We
have derived basic equations concerning the relativistic diffusion
with friction in a gravitational background (adding friction to
the diffusion equations of Franchi and Le Jan \cite{lejan}) . Such
equations can describe the dynamics of particles around stars and
black holes in the presence of a gas of some other particles. The
gravitational effects must be sufficiently strong to be detectable
by experiments. The most impressive data apply to CMBR spectrum.
Then, the diffusion of photons scattered on charged particles in
the intergalactic space can distort the CMBR spectrum
\cite{brink}. An application of the relativistic diffusion to the
CMBR spectrum distortion (Sunyaev-Zeldovich effect) is discussed
in \cite{habampa}. The effect of the gravitational field studied
in sec.7 is rather weak to be detected in a near future.
Nevertheless, from the derivation of the (Kompaneetz) diffusion
equation \cite{rybicki} it can be seen that if the quantum
electrodynamics in a background metric is considered then the
resulting diffusion process should be the one perturbing the
geodesic motion.
  The relativistic diffusion can be a useful tool
for a selection of a relativistic approximation to multiparticle
interactions determined by the form of the equilibrium resulting
from the relativistic quantum field theory at finite temperature.
We discussed as an example a diffusion  in de Sitter space. We
obtained a surprising relation between the diffusion on the de
Sitter space and the diffusion on the flat space in a heat bath of
non-zero temperature. There is a well-known relation between the
particle temperature in the de Sitter space and the radius of this
space \cite{hawk}. Such a relation results from a quantum theory
and it is different from the one derived in this paper.

 The main aim of this paper is a
discussion of the friction terms which lead to an equilibrium.
  It is not clear what is the physical  meaning of the diffusion
 in the proper time because in contradistinction to the deterministic dynamics in random dynamics
the proper time associated with an observer moving with  a
particle is a random variable. The transport equation (sec.5)
expressed in the laboratory time has a clear physical meaning. We
discuss the transport equation in more detail in \cite{habajmp}.
In particular, we show that the diffusion in the laboratory time
can be obtained from the diffusion in the proper time by an
integration over the proper time in a similar way as  this is done
in Feynman's relativistic dynamics \cite{fey} (for a discussion of
the relation between the proper time dynamics and laboratory time
dynamics see also \cite{dun}).

\section{Appendix A:Stochastic equations}
In this Appendix we explain the relation between the diffusion
equation (24) and the stochastic equations (29)-(31) (see
\cite{ikeda} for a complete theory) . The main observation is that
the solution at time $\tau$ of the stochastic differential
equations (assuming it is unique) is causal,i.e., depends on the
initial conditions and values of the Brownian motion ${\bf b}(s)$
at $s\leq \tau$. As a consequence (of the Markov property) eq.(26)
determines a semigroup $T_{\tau}\phi=\phi_{\tau}$. In such a case
in order to prove that $\phi_{\tau}$ of eq.(26) is the solution of
eq.(24) it is sufficient to check that the initial conditions are
the same and the generators coincide. The initial condition for
eq.(24) follows from the choice
 of the initial conditions for the stochastic equations. Then, the
calculation of the generator is reduced to the solution of the
stochastic equations at arbitrarily small time and the calculation
of $d\phi$. Let us consider eqs.(29)-(30) for $x^{\mu}$ and assume
eq.(31) in a  general form
\begin{equation}
dp_{j}=B_{j}d\tau+mc\kappa e_{j}^{a}db_{a},
\end{equation}
where $e_{j}^{a}$ are defined in eqs.(27)-(28) (together with the
$\gamma_{jk}$ of eq.(9) and $g_{jk}=B^{-2}\delta_{jk}$ for
isotropic coordinates). As a next step we need to calculate (for a
small $\tau$)
\begin{equation}
\phi(x_{\tau},{\bf p}_{\tau})-\phi(x,{\bf
p})\simeq\phi(x+\triangle x,{\bf p}+\triangle {\bf p})-\phi(x,{\bf
p}),
\end{equation}
where from eqs.(29)-(30) $\triangle x^{0}=\frac{1}{m}A\omega
\triangle \tau$, $\triangle {\bf x}=-\frac{1}{m}B^{2}{\bf
p}\triangle \tau$ and from eq.(96)
\begin{equation}
\triangle p_{j}=B_{j}\triangle \tau+mc\kappa
e_{j}^{a}(b_{a}(\triangle\tau)-b_{a}(0)),
\end{equation}
here $b_{a}(0)=0$. We expand eq.(97) in a Taylor series in
$\triangle x^{\mu}$ and $\triangle p_{j}$ (or what is the same in
$\triangle \tau$ and $ b_{a}(\triangle \tau)$). After the
expansion we calculate the expectation values involving the
Brownian motion (see the expectation value below eq.(31); we use
$E[b_{a}]=0 $ and $E[b_{a}(\triangle \tau)b_{c}(\triangle
\tau)]=\delta_{ac}\triangle \tau$). We must take into account
terms till the second order in $\triangle p$ which give
\begin{equation} E[\triangle p_{j}\triangle
p_{k}]=m^{2}c^{2}\kappa^{2}e_{j}^{a}e_{k}^{c}E[b_{a}(\triangle\tau)b_{c}(\triangle\tau)]
=m^{2}c^{2}\kappa^{2}e_{j}^{a}e_{k}^{a}\triangle\tau
=m^{2}c^{2}\kappa^{2}\gamma_{jk}\triangle \tau.\end{equation} Now,
collecting the terms of order $\triangle \tau$ we obtain for
$\tau=\triangle \tau$
\begin{equation}\begin{array}{l}
\phi_{\tau}-\phi=\frac{1}{m}A\omega\frac{\partial}{\partial
x^{0}}\phi \triangle\tau
-\frac{1}{m}B^{2}p_{j}\frac{\partial}{\partial x^{j}}\phi\triangle
\tau \cr+B_{j}\frac{\partial \phi}{\partial p_{j}}\triangle \tau
+m^{2}c^{2}\kappa^{2}\frac{1}{2}\gamma_{jk}\frac{\partial}{\partial
p_{j}}\frac{\partial}{\partial p_{k}}\phi\triangle \tau.
\end{array}\end{equation}
Dividing by $\triangle\tau$ and taking the limit
$\triangle\tau\rightarrow 0$ we obtain eq.(24) with
\begin{displaymath}
B_{j}=\frac{1}{2m}{\bf p}^{2}\frac{\partial B^{2}}{\partial x^{j}}
-\frac{1}{2m}\omega^{2}\frac{\partial \ln A}{\partial
x^{j}}+\frac{3}{2}\kappa^{2}p_{j}.\end{displaymath}

\section{Appendix B:Time evolution of the energy}
We calculate the change of the energy
\begin{equation}
dp_{0}=d(A^{-1}\omega)=-\omega A^{-2}\partial_{j}Adx^{j}
-A^{-1}\omega^{-1}B\partial_{j}B{\bf p}^{2}dx^{j}+
A^{-1}B^{2}\omega^{-1}p_{j}\circ dp_{j}.
\end{equation}
Here, the circle denotes the Stratonovitch differential
\cite{ikeda}
\begin{equation}
h\circ dg=hdg+\frac{1}{2}dhdg
\end{equation}
The correction term to the Stratonovitch differential is
\begin{equation}
\frac{1}{2}d(\omega^{-1}p_{j})dp_{j}=\frac{3}{2}m^{2}c^{2}\kappa^{2}B^{-2}\omega^{-1}d\tau.
\end{equation}
 After an insertion of the differentials (29)-(30) and (43) on the
rhs of eq.(101) the deterministic ($\kappa$ independent terms)
cancel identically (as they should because $p_{0}$ is a constant
of motion when the stochastic perturbation is absent). There
remains
\begin{equation}\begin{array}{l}
 dp_{0}=\frac{3\kappa^{2}}{2}A^{-1}\omega^{-1}B^{2}{\bf p}^{2}d\tau
+\frac{3}{2}m^{2}c^{2}\kappa^{2}A^{-1}\omega^{-1}d\tau+\frac{\kappa^{2}}{2}A^{-2}B^{2}c\beta
f^{\prime}{\bf p}^{2}d\tau\cr
 +\kappa A^{-1}B{\bf p} d{\bf b}.\end{array}
\end{equation}
This is eq.(55) (after elementary transformations).

A proof of the formula (52) for the angular momentum is much
simpler. We have\begin{equation} d{\bf L}=d{\bf x}\times {\bf
p}+{\bf x}\times d{\bf p}={\bf x}\times d{\bf p}
\end{equation}We assume that the functions $A$ and $B$ in the
isotropic metric are spherically symmetric. Then, in $d{\bf p}$
(defined in eq.(43)) the terms independent of $\kappa$ are
parallel to ${\bf x}$ because gradients of $A$ and $B$ are
parallel to ${\bf x}$. The $\kappa$-dependent terms in (105) give
eq.(45).


\begin{thebibliography}{99}
\bibitem{schay} G.Schay,PhD thesis,Princeton University,1961
\bibitem{dudley} R.Dudley, Arkiv for Matematik,{\bf 6},241(1965)



\bibitem{lejan}J. Franchi and Y. Le Jan, Commun.Pure Appl.Math.{\bf 60},187(2007)
\bibitem{fran}J. Franchi, Commun.Math.Phys.{\bf 290},523(2009)
\bibitem{franha}J. Angst and J. Franchi, Journ.Math.Phys.{\bf 48},083103(2007)
\bibitem{bai}I.Bailleul,Prob.Theory Relat. Fields{\bf 141},283(2008)

\bibitem{bai2}I. Bailleul, arXiv:0810.5662
\bibitem{debbasch}F. Debbasch, Journ. Math.Phys.{\bf
45},2744(2004)
 \bibitem{haba}Z. Haba, Phys.Rev.{\bf E79},021128(2009)
\bibitem{habampa}Z. Haba, Mod.Phys.Lett.{\bf  24A},3193(2009)
\bibitem{habajpa}Z. Haba, Journ.Phys.{\bf A42},445401(2009)
\bibitem{chev2}C. Chevalier and F. Debbasch, Journ.Math.Phys.{\bf
49},043303(2008)
\bibitem{hara}S. W. Hawking, Commun.Math.Phys.{\bf
43},199(1975)
\bibitem{chev1}C. Chevalier and F. Debbasch, AIP Conf.Proc.{\bf
913},42(2007)

\bibitem{han}J. Dunkel and P. H\"anggi,Phys.Rep.{\bf 471},1(2009)
\bibitem{in}R. D'Inverno, Introducing Einstein's
Relativity, Clarendon Press, Oxford,1996
\bibitem{klein} P. Collas and D. Klein, Gen.Rel.Grav.{\bf
39},737(2007)
\bibitem{ikeda} N. Ikeda and S. Watanabe, Stochastic
Differential Equations and Diffusion Processes,North Holland,1981



\bibitem{jut} F. J\"uttner, Ann.Phys.(Leipzig){\bf 34},856(1911)
\bibitem{habajmp}Z.Haba,arXiv:0911.3126
\bibitem{hasm}R.Z.Khasminski, Stochastic Stability of Differential
Equations, Sijthoff and Noordhoff,1980
\bibitem{las}A. Lasota and G. Mackey, Probabilistic Properties of Deterministic Systems,
Cambridge University Press,Cambridge ,1985
\bibitem{heavy}R.Rapp and H. van Hees, arXiv:0803.0901

\bibitem{rybicki}G.B. Rybicki and A.P. Lightman,

Radiative Processes in Astrophysics,Wiley,1979
\bibitem{yor}M.Yor, Z.Wahr.Verw.Gebiete,{\bf 53},71(1980)

H. Matsumoto and M. Yor, Probability Surveys,{\bf 2},312(2005)


\bibitem{hawk}G.W. Gibbons and S.W. Hawking  , Phys.Rev.{\bf
D15},2738(1977)




\bibitem{brink}M. Birkshaw, Phys.Rep.{\bf 310}97(1999)
\bibitem{fey}
R.P. Feynman, Phys.Rev.{\bf 80},440(1950),{\bf 84},108(1951)

\bibitem{dun}J. Dunkel, P. H\"anggi and S. Weber, Phys.Rev.{\bf E
79},010101(R)(2009)

\end{thebibliography}
\end{document}